hep-lat/9401020 18 Jan 94

# The scalar and tensor glueballs in the valence approximation*

H. Chen, J. Sexton†, A. Vaccarino and D. Weingarten

IBM Research, P.O. Box 218, Yorktown Heights, NY 10598

We evaluate the infinite volume, continuum limit of $0^{++}$ and $2^{++}$ glueball masses in the valence approximation. We find $m_{0^{++}} = 1740 \pm 71$ MeV and $m_{2^{++}} = 2359 \pm 128$ MeV, consistent with the interpretation of $f_0(1710)$ as the lightest scalar glueball.

Encouraging results obtained recently for light hadron masses in the valence approximation [1] have prompted us to investigate the glueball spectrum. For both $0^{++}$ and $2^{++}$ states, a quark-antiquark system has odd orbital angular momentum. The resulting angular momentum barrier should tend to suppress the quark-antiquark annihilation required for mixing with glue states. An argument [2] based on the observed near degeneracy of corresponding pairs of isovector and isoscalar mesons composed of a quark and an antiquark with nonzero orbital angular momentum supports this expectation. Thus we believe there is a reasonable chance that our results provide fairly reliable predictions for the real world.

We use square loop glueball operators constructed from smeared links. Smearing eliminates some of the high frequency noise and increases the projection to the ground state [3]. Although different methods have been proposed and used in the past, most of them consist of defining smeared $SU(3)$ link matrices by adding and multiplying together neighboring links in a gauge invariant way [4–7]. Part of our work uses the operators proposed in Ref. [7], in which the smearing of Ref. [4] was used to construct arbitrarily large $N_L \times N_L$ loops. In Ref. [7] it was also found that this method is similar in performance to the smearing proposed by Tepper [5] and by DeGrand [6]. Our operators are then characterized by a the smearing strength $\epsilon$, smearing level $N_S$ and loop size $N_L$ [7].

We also performed a calculation using Coulomb gauge smeared operators. We define smeared link variables $U_i^s(x)$ for $s \geq 1$ by the average

$$U_i^s(x) = \frac{1}{(s+1)^2} \sum_{0 \leq p,q \leq s} U_i(x + p\hat{j} + q\hat{k}), \quad (1)$$

where $\hat{j}$ and $\hat{k}$ are the two space directions orthogonal to $\hat{i}$. Here the smeared $U_i^s(x)$ is not projected to $SU(3)$, as is done in the case of the gauge invariant operators. The product of $s$ sequential $U_i^s(y)$

$$V_i^s(x) = U_i^s(x) \ldots U_i^s[x + (s-1)\hat{i}] \quad (2)$$

is then used to construct $s \times s$ loops. Therefore each operator is uniquely specified by the parameter $s$.

At larger $\beta$ we find the gauge invariant operators couple more efficiently to glueball ground states than do the Coulomb gauge operators. Even at smaller $\beta$, however, the required gauge fixing makes Coulomb gauge operators computationally more expensive.

Two vacuum subtracted glueball operators $O_s$ and $O_{s'}$, with the same quantum numbers $J^{PC}$ and different smearing parameters $r$ and $r'$, generate a propagator $C_{rr'}(t)$ that at large time separations approaches the asymptotic form

$$Z_r Z_{r'}^* \{exp(-mt) + exp[-m(T-t)]\}. \quad (3)$$

Here $T$ is the lattice temporal period, $m$ is the mass of the lightest state, and $Z_r$ is the projection $\langle \Omega | O_r | J^{PC} \rangle$. For Coulomb gauge smearing, $r$ is the size parameter $s$, while for gauge invariant smearing $r$ represents the triple $(\epsilon, N_S, N_L)$. The Coulomb gauge data was analyzed by fitting the propagators obtained from different operators to Eq. (3) all at once over a range of several time slices. All the fits were performed using the full correlation matrix for statistical errors. For the gauge invariant data we fitted only diagonal propagartors $C_{rr}(t)$ to Eq. (3). The most probable combined mass was then found using the statistical correlation matrix among the fitted masses for different $r$. The required correlation matrix, and the statistical errors in all fits, were found by the bootstrap method.

---

*Talk presented by A. Vaccarino

†Permanent Address: Department of Mathematics, Trinity College, Dublin 2, Republic of Ireland

Table 1
Summary of our calculations and results.

| $\beta$ | Lattice | Count | Skip | $am_{0^{++}}$ | $am_{2^{++}}$ |
|---|---|---|---|---|---|
| 5.70 | $16^3 \times 24$ | 6,050 | 400+ 0 (1) | 0.964(42) | — |
|  |  | 8,094 | 10+40 (8) | 0.983(40) | — |
| 5.83 | $20^3 \times 30$ | 4,002 | 400+ 0 (1) | 0.858(43) | — |
| 5.93 | $24^3 \times 36$ | 4,004 | 400+ 0 (1) | 0.811(33) | 1.144(107) |
|  | $16^3 \times 24$ | 30,640 | 5+20 (16) | 0.786(12) | 1.266(36) |
|  | $12^3 \times 24$ | 48,278 | 5+20 (16) | 0.771(10) | 1.202(26) |
| 6.17 | $32^2 \times 30 \times 40$ | 2,005 | 400+ 0 (1) | 0.489(31) | 0.816(119) |
|  | $24^3 \times 36$ | 31,150 | 5+20 (16) | 0.582(10) | 0.828(26) |
| 6.40 | $32^2 \times 30 \times 40$ | 2,002 | 400+ 0 (1) | 0.415(43) | 0.504(61) |
|  |  | 25,440 | 5+20 (16) | 0.433(11) | 0.636(23) |

Table 1 summarizes our calculations and results. For each $\beta$ the first row refers to Coulomb gauge operators and the second, if present, to gauge invariant operators. The smearing strength parameter $\epsilon$ was set to 0.25 at $\beta = 5.7$, and 1.0 in all other cases. The table also lists the number of measurements and the number of $SU(3)$ Monte Carlo updates between each measurement in the format heat bath sweeps + microcanonical sweeps [8]. Successive measurements in all cases appeared to be nearly independent statistically. In some cases we found a slight increase in error bars if successive propagators were binned together to produce a ensemble with fewer members and smaller correlations. In parentheses we indicate the bin sizes, generally taken much larger than needed, used in our final evaluation of error bars. The two different smearing methods generally produced statistically consistent mass results. We found, however, some discrepancies at the two largest $\beta$, presumably due to the small Coulomb gauge ensembles yielding unrealiable estimates of the size of their errors. Our masses from the largest ensembles are all consistent with recent calculation by other groups at similar values of $\beta$ [9,10]. Figs. 1 and 2 show typical effective mass plots. A plateau is quite clear in the $0^{++}$ data, and present, but somewhat more ambiguous, for the $2^{++}$.

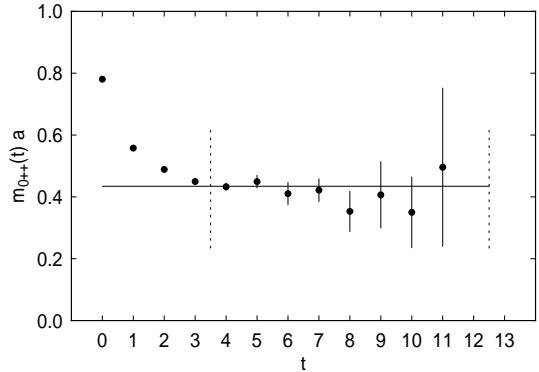

Figure 1. Effective mass plot for the $0^{++}$ glueball at $\beta = 6.4$ for the operator with smearing parameters $(\epsilon, N_S, N_L) = (1.0, 8, 9)$. We also show the fitted mass and the fitting range.

The distance of our masses from the infinite volume limit can be estimated using Lüscher's formula [11]

$$m(z) = m(\infty)\{1 - g\frac{exp(-\sqrt{3}z/2)}{z}\}, \qquad (4)$$

where $z = mL$ and $L$ is the lattice period. From our data at $\beta = 5.93$, we found $g_{0^{++}} = 682 \pm 434$

and $g_{2^{++}} = (1.72\pm0.85)\times10^5$, consistent with the estimate of Ref. [12]. The gauge invariant data at the three largest $\beta$, which we used for the continuum limit extrapolation, all have $z_{0^{++}} > 12$ and $z_{2^{++}} > 19$, which imply masses within 0.5 % of their infinite volume limits. This error is significantly smaller than the statistical uncertainty in each mass.

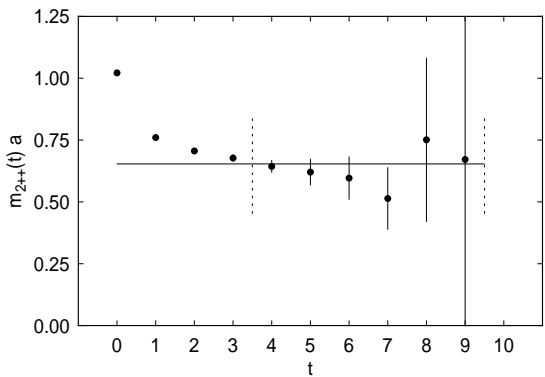

Figure 2. Same as Fig. 1, but for the $2^{++}$ glueball.

To set the lattice scale $a$ we used $\Lambda_{\overline{MS}}^{(0)}$ from Ref. [1]. Both $m_\rho a$ and $m_\phi a$ were found to obey asymptotic scaling within statistical errors above $\beta$ of 5.7. The continuum limit of $(\Lambda_{\overline{MS}}^{(0)}a)/(m_\phi a)$ gives $\Lambda_{\overline{MS}}^{(0)} = 243.7 \pm 6.8$ MeV, consistent with independent estimates [13,14]. Since the gauge part of the QCD action has $O(a^2)$ corrections, we extrapolated mass ratios linearly in $\left(a\Lambda_{\overline{MS}}^{(0)}\right)^2$, and obtained the continuum limits $m_{0^{++}}/\Lambda_{\overline{MS}}^{(0)} = 7.14 \pm 0.21$ and $m_{2^{++}}/\Lambda_{\overline{MS}}^{(0)} = 9.68 \pm 0.45$. Extrapolations are shown in Fig. 3. Combining the errors in ratios and in $\Lambda_{\overline{MS}}^{(0)}$, we obtained $m_{0^{++}} = 1740\pm71$ MeV and $m_{2^{++}} = 2359\pm128$ MeV. The mass of the observed $f_0(1710)$ is close to our estimate, making it our preferred glueball candidate. The $2^{++}$ is close to at least two well established resonances, $f_2(2300)$ and $f_2(2340)$, which can not be distinguished within our statistical uncertainties.

We would like to thank Frank Butler for writing some of the analysis software which we used, and Mike Cassera, Molly Elliott, Dave George, Chi Chai Huang and Ed Nowicki for their work on GF11.

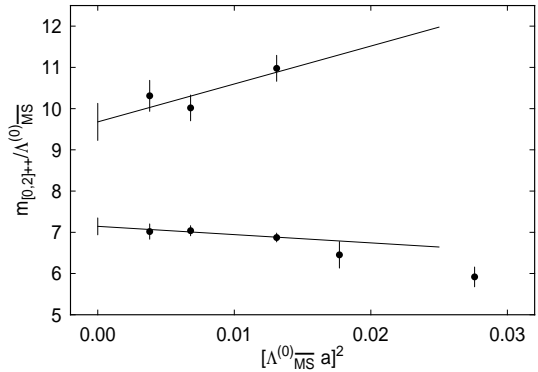

Figure 3. The continuum limit extrapolation for both the $0^{++}$ and $2^{++}$ glueball.